\documentclass[11pt]{article}
\pdfoutput=1

\usepackage[utf8]{inputenc}
\usepackage{multirow}
\usepackage{amsmath, amsfonts, amssymb}
\usepackage{comment}
\usepackage{graphicx}
\usepackage{psfrag}
\usepackage{amsthm}
\usepackage[usenames,dvipsnames,svgnames,table]{xcolor}
\usepackage{enumerate}
\usepackage{arydshln}
\usepackage{soul}
 \usepackage{slashed}
\usepackage{subfig}
\usepackage{mathrsfs}
 \usepackage{a4wide}
  \usepackage{tikz}
  \usepackage{tikz-cd}
  \usetikzlibrary{shapes.geometric}
  \usepackage{tcolorbox}

  \usepackage{color}
  \definecolor{dark-gray}{gray}{0.20}
  \definecolor{gray}{gray}{0.30}
  \definecolor{light-gray}{gray}{0.80}
  \definecolor{dark-red}{rgb}{0.7,0,0}
  \definecolor{dark-green}{rgb}{0.1,0.4,0}
  \definecolor{dark-blue}{rgb}{0.3,0.3,0.7}
  \definecolor{light-blue}{rgb}{0.8,0.8,1}
      \definecolor{swamp}{RGB}{240, 199, 197}

  \usepackage{pifont}

\usepackage{newunicodechar} 
\newunicodechar{ỳ}{\`y}

\usepackage{setspace}

\newcommand{\be}{\begin{equation}}
\newcommand{\ee}{\end{equation}}

\def\be{\begin{equation}}
\def\ee{\end{equation}}
\def\bea{\begin{eqnarray}}
\def\eea{\end{eqnarray}}

\newcommand{\beq}{\begin{equation}}  \newcommand{\eeq}{\end{equation}}
\newcommand{\bal}{\begin{aligned}}   \newcommand{\eal}{\end{aligned}}
\def\beqa{\begin{eqnarray}}
\def\eeqa{\end{eqnarray}}

\def\simleq{\; \raise0.3ex\hbox{$<$\kern-0.75em
      \raise-1.1ex\hbox{$\sim$}}\; }
   \def\simgeq{\; \raise0.3ex\hbox{$>$\kern-0.75em
      \raise-1.1ex\hbox{$\sim$}}\; }

\numberwithin{equation}{section}

\usepackage{jheppub}
\usepackage{hyperref}
\usepackage{cleveref}

\hypersetup{
	colorlinks=true,
	linkcolor=dark-blue,
	citecolor=dark-red,
	urlcolor=dark-green,
	linktoc=page
}

\theoremstyle{remark}

\crefname{appendix}{Appendix}{Appendices}

\title{\centering The Dark Dimension and the Swampland}

\author{Miguel Montero$^1$,} 
\author{Cumrun Vafa$^1$ and}
\author{Irene Valenzuela$^{2,3}$}
\affiliation{$^1$Jefferson Physical Laboratory, Harvard University, Cambridge, MA 02138, USA}
\affiliation{$^2$Instituto de F\'{i}sica Te\'{o}rica UAM-CSIC and Departamento de F\'{i}sica Te\'{o}rica, Universidad Aut\'{o}noma de Madrid, Cantoblanco, 28049 Madrid, Spain}
\affiliation{$^3$CERN, Theoretical Physics Department, 1211 Meyrin, Switzerland}

\emailAdd{mmontero@g.harvard.edu}
\emailAdd{vafa@g.harvard.edu}
\emailAdd{irene.valenzuela@cern.ch}

\abstract{Motivated by principles from the Swampland program, which characterize requirements for a consistent UV completion of quantum gravity, combined with observational data, we are led to a unique corner of the quantum gravity landscape.  In particular, using the Distance/Duality conjecture and the smallness of dark energy, we predict the existence of a light tower of states and a unique extra mesoscopic dimension of length $l\sim \Lambda^{-\frac{1}{4}}\sim  10^{-6}\, m$, with extra massless fermions propagating on it.  This automatically leads to a candidate for a tower of sterile neutrinos, and an associated active neutrino mass scale $m_{\nu}\sim \langle H\rangle^2\,  \Lambda^{-\frac{1}{12}}M_{pl}^{-\frac{2}{3}}$. 
Moreover,  assuming the mechanism for stabilization of this dark dimension leads to similar masses for active and sterile neutrinos we are led to the prediction of a Higgs vev $\langle H\rangle \sim \Lambda^{\frac{1}{6}}M_{pl}^{\frac{1}{3}}$.  Another prediction of the scenario is a species scale ${\hat M} \sim  \Lambda^ {\frac{1}{12}}M_{pl}^{\frac{2}{3}}\sim 10^{9}-10^{10} GeV$, corresponding to the higher-dimensional Planck scale. This energy scale may be related to the resolution of the instability of the Higgs effective potential present at a scale of $\sim 10^{11}\, GeV$. We also speculate about the interplay between this energy scale and the GZK limit on ultra-high energy cosmic rays.}

\begin{document}
\hypersetup{pageanchor=false}
\makeatletter
\let\old@fpheader\@fpheader

\makeatother

\maketitle

\hypersetup{
    pdftitle={A cool title},
    pdfauthor={Cody Long, Miguel Montero, Cumrun Vafa, Irene Valenzuela},
    pdfsubject={Finiteness of the string Landscape, Calabi-Yau}
}

\emailAdd{mmontero@g.harvard.edu}
\emailAdd{irene.valenzuela@uam.es}
\emailAdd{vafa@g.harvard.edu}
\emailAdd{mmontero@g.harvard.edu}

\newcommand{\remove}[1]{\textcolor{red}{\sout{#1}}}

\newpage

\section{Introduction}
\label{sec:intro}
Naturalness issues, including the hierarchy problem as well as the smallness of the dark energy, continue to be major puzzles for theoretical physics.
The usual notion of naturalness in quantum field theories assumes that including quantum gravity in the mix does not lead to a dramatic modification of this view. However the Swampland program \cite{Vafa:2005ui} (see \cite{Brennan:2017rbf,Palti:2019pca,vanBeest:2021lhn} for reviews), motivated from lessons learned from string theory, challenges this perspective and has led to a dramatic reformulation of the meaning of naturalness in the context of quantum gravitational theories and to strong restrictions on EFT's consistent with quantum gravity.

In this note we point out that some of the basic ideas of the Swampland program, combined with certain observational data, naturally lead us to a particular corner of the string landscape, corresponding to an asymptotic region of the field space. The cosmological hierarchy, namely the smallness of dark energy in Planck units, is the main driver of this. By Swampland considerations, this asymptotic region is necessarily accompanied by a light tower of states which, combined with observational data, lead to the prediction of a single extra mesoscopic dimension of length $l\sim  \lambda \Lambda ^{-1/4}\sim 1\, 
\mu m$, where $\Lambda$ is the cosmological constant and we estimate $\lambda \sim 10^{-1}-10^{-3}$. Moreover combined with other observational data this leads to the identification of a tower of sterile neutrinos with mass scale $m_{\nu_s} \sim \lambda^{-1}\Lambda^{1/4}\sim 1\, eV$, and the emergence of a new scale ${\hat M}\sim \lambda^{-1/3} \Lambda^{1/12}M_{pl}^{2/3}\sim 10^{9}-10^{10}GeV$, where new tower of matter states charged under Standard Model gauge fields emerge and gravity becomes strongly coupled.  This is tantalizingly close to the scale of $10^{11}GeV$ where the Higgs potential is believed to develop instability; in our setup the tower charged under Standard Model gauge fields will lead to its modification and presumably also stability (assuming that is also related to the scale of supersymmetry breaking) as we will explain.  Absence of strong mass hierarchies in the neutrino sector, and more specifically  requiring that active neutrinos and sterile neutrinos not to have very different masses leads to $m_H \sim y^{-1}\lambda^{-2/3} \Lambda^{1/6} M_{pl}^{1/3}$ (with $y\sim 10^{-2}-10^{-3}$ denoting the coupling of Standard Model fields to KK modes).  Thus not only the hierarchy, the neutrino mass scale and the dark energy have been connected together through consistency principles of quantum gravity, but we also have a prediction of a new physical scale ${\hat M}\sim 10^{9}-10^{10}GeV$.

 The notion of bringing down the Planck scale/string scale down to near the weak scale (by considering large extra dimensions) was considered in \cite{Antoniadis:1990ew,Arkani-Hamed:1998jmv,Dienes:1998sb,Randall:1999ee,Randall:1999vf}.  However the motivation and the results in these works are rather different from ours. In particular the aim in these works was to solve the electroweak hierarchy problem through unification of weak scale with a higher dimensional Planck scale in the TeV range.  This in particular originally led to $n\geq 2$ large extra dimensions in the scenario studied in \cite{Arkani-Hamed:1998jmv} (and current bounds are much more stringent, leading to $n\geq 4$ for typical extra dimensions of size less than $10^{-12}m$ \cite{Fermi-LAT:2012zxd}). We are instead motivated by the existence of small dark energy and other experimental observations to predict exactly one mesoscopic dimension in the micron range. In other words, we tie the existence of a large extra dimension to the \emph{cosmological} hierarchy problem rather than the electroweak one.   However, interestingly, the weak scale hierarchy also ends up being related to the dark energy being small, as we will discuss later in this paper.
 While there has been other works suggesting relations between dark energy and large extra dimensions (see e.g.\cite{Aghababaie:2003wz,Sundrum:2003jq,Dupays:2013nm}) this relation has not been connected to the requirement for consistency of quantum gravity as predicted by the Swampland criteria that we explore in this work.

The organization of this paper is as follows: In section \ref{sec:species-2} we discuss some of the main criteria that we will use from the Swampland program. In section \ref{sec:dd-3} we use these to explain why this leads to a single dark dimension of micron length scale and the emergence of $10^9-10^{10}$ GeV as a new fundamental  scale in physics. In section \ref{sec:fth-4} we discuss possible realization of this scenario in string theory via F-theory compactifications.  In section \ref{sec:pheno-5} we discuss further phenomenological predictions of this model.

\section{Some Swampland Principles}\label{sec:species-2}
One of the main, well-tested Swampland conditions is the distance/duality conjecture \cite{Ooguri:2006in}, which states that at large distance in field space $\phi$ we get an exponentially light tower of states with mass $m\sim \exp(-\alpha \phi)$ where $\alpha \sim O(1)$ in Planck units. This light tower of states is weakly coupled and leads to a dual description of the theory.  Moreover, due to the tower of light states, there will be a maximal energy cut-off at which any local QFT description breaks down and quantum gravity becomes important. This is the so-called ``species scale'' \cite{Dvali:2007hz,Dvali:2007wp}.  Thus to each tower we can associate two mass scales:  $m$, which is the mass scale of states in the tower and $\Lambda_{sp}$ which is the scale local QFT description breaks down.  The value of this quantum gravity cut-off depends on the nature of the tower.
 
 A priori, the tower of states could have any microscopic origin. However, only two cases have ever been encountered so far in all known asymptotic limits of string theory compactifications.
\begin{itemize}
    \item A tower of string excitation modes. Due to the exponential degeneracy of states, the species scale is near the string scale $\Lambda_{sp}\sim m$, where one hits the first higher spin state. The theory is still weakly coupled to gravity at that scale, but local QFT breaks down due to the presence of higher spin states.
    \item A Kaluza-Klein tower signaling decompactification. One or more extra dimensions open up at the scale $m$ of the tower, and the physics can still be described by a QFT in higher dimensions until we reach the species scale, which in this case corresponds to the Planck scale of the higher dimensional theory. This is given (for 4 macroscopic dimensions) by
    \beq
   \hat{M}\equiv  \Lambda_{sp}=M_{pl,n}=m^{\frac{n}{n+2}}M_{pl}^{\frac{2}{2+n}}
   \label{MPn}
   \eeq
   where $n$ is the number of effective dimensions decompactifying. The physics becomes strongly coupled to gravity beyond this scale, so that it cannot be described by effective field theory.
    
\end{itemize}
Based on all the known examples from string theory it was conjecture in \cite{Lee:2019wij,Lee:2019xtm} that these are indeed the two only possibilities in quantum gravity (the ``Emergent String Conjecture"). 

Distance conjecture has also been studied in the context of AdS or dS vacua and it has been argued in \cite{Lust:2019zwm} that the natural distance scale in field space is proportional to $ \rm{log}( 1/|\Lambda|)$ as $\Lambda$ itself can be viewed in quasi-static case as being parameterized by a field leading to a generalized notion of distance. This leads to a tower of light states 
\begin{equation} m\sim |\Lambda|^{\alpha}. \label{alpha} \end{equation}
Even though the Swampland argument is the same in the dS or AdS cases, the AdS case is supported by a far larger amount of evidence in string theory. Nevertheless, we will assume this continues to hold in dS as well, as was originally argued.  Note that the dS distance conjecture is in a sense a `solution' to the cosmological constant problem.  Because the above relation can be reinterpreted as 
\begin{equation}
    \Lambda \sim m^{1\over \alpha}
\end{equation}
This in particular implies that as $m\rightarrow 0$, we should have $\Lambda\rightarrow 0$.  In particular the massive states naively seem not to contribute to the cosmological constant.  From EFT we may have expected to have
\begin{equation}
    \Lambda \sim \Lambda_0 +A \ m^{1\over \alpha}
\end{equation}
where $\Lambda_0$ would be the contribution of all the heavy modes.  We are seeing that the distance conjecture amounts to saying that $\Lambda_0=0$.  

To gain insight, and more confidence, in the dS version of the distance conjecture, let us check how this works in the few cases that we are able to check in string theory.
Consider the limit of vanishing string coupling $g=0$.  In this case the tower of light states is the string states which in the strict limit lead to $m=0$.  So in this case we should get the vanishing of the cosmological constant. Indeed we know this is true in all the string theories, as at the tree level cosmological constant is zero.  This is not because of supersymmetry or any magical boson/fermion cancellation.  It is even true in bosonic string theory.
One way to see how this arises is to note that the contribution of sphere amplitude is always divided by the volume of the symmetry group which at genus 0 is the conformal group $SL(2,C))$, and so we end up with
\begin{equation}
    \Lambda_0={K\over {\rm Vol}(SL(2,C))}=0,
\end{equation}
since the ${\rm Vol}(SL(2,C)=\infty$.  So the conformal invariance of string theory is responsible for this magical cancellation! 

This magic continues at higher loops as well.  
Consider the example of the non-supersymmetric $O(16)\times O(16)$ heterotic strings \cite{Alvarez-Gaume:1986ghj,Dixon:1986iz}.  In this case the leading non-vanishing contribution to the energy at weak string coupling arises from one loop string effect, leading to $V\sim m^{10}$ where $m$ is the string scale.  Indeed at weak coupling, string excitations form a light tower. Again it seems surprising that the more massive string excitations, or non-perturbative massive branes (such as the NS 5-brane) do not overwhelm the light state contributions. Again the remenant of conformal symmetry which at 1-loop is the $SL(2,Z)$ symmetry known as modular invariance is what shuts off the contribution of higher states to the one-loop vacuum energy energy. 
In particular, the contributions of massive string excitations of masses $m_i$ to the one-loop vacuum energy, which are expected to be of the form
\begin{equation}V_{\text{1-loop}}\sim- \sum_{i}(-1)^{F_i} \int_{\Lambda_{UV}^{-2}}^{\infty} \frac{ds}{s^6}\  {\rm exp}\left(-\frac{m_i^2s}{2}\right),\end{equation}
are cutoff at the Schwinger parameter $s\sim m_s^{-2}$ due to modular invariance \cite{polchinski1998string}. Here $(-1)^{F_i}$ is the fermion number of each particle species in the sum. This makes the high energy cutoff scale $\Lambda_{UV}$ (the species bound)  equal to the string scale $m_s$.  Note that the reason for this cutoff is not that at higher mass scales string perturbation breaks down, because this does not happen until we get to very high masses $m_s/g$.  So it is magical and surprising from the EFT point of view that the string states with masses in the range
\begin{equation}
    O(1)\cdot m_s<m<{m_s\over g}
\end{equation}do not seem to play a major role in determining the vacuum energy.  We can view the modular invariance shutting off of the more massive stringy excitations as trying to avoid redundancy of counting the same states more than once.  Similarly the fact that at weak string coupling the non-perturbative branes do not contribute to vacuume energy can be attributed to the fact that they can be viewed as composite objects made of light strings and it would be redundant to add the extra contributions. A field-theoretic analog of this effect is the fact that monopoles do not run in loops at weak coupling, even if they are stable particles.  In other words both supressions may be viewed as trying to avoid overcounting the contribution of states and that the light string states constitute the fundamental entities in the weak coupling limit of strings.
In gravity, this feature dictated by the distance/duality conjecture is the restatement that the light tower of states give rise to a complete effective dual description of the theory.  The fact that the energy is proportional to some power of $m$, the mass scale of the light tower, is also consistent with the emergence proposal, which attempts to explain how large distances in field space, small gauge couplings and related physical quantities arise from integrating out the light tower \cite{Heidenreich:2017sim,Grimm:2018ohb,Heidenreich:2018kpg}.

In the dS or quasi-dS case, it was further argued in \cite{Lust:2019zwm} that $\alpha \leq\frac12$ to be  consistent with the Higuchi bound \cite{Higuchi:1986py}.
Indeed in some stringy examples there are stronger contributions to $V$ coming from tree level leading to $V\sim m^2$.
We now argue that $\frac{1}{d}$ is the lower bound on $\alpha$:
\begin{equation}\frac{1}{d} \leq \alpha \leq \frac12.\label{mainb}\end{equation}

  Given that the tower of light states has a mass scale $m$, one would expect at the very least a one-loop energy contribution from the light tower proportional to $V\sim m^d$ setting a scale for dark energy\footnote{This observation has already been used in \cite{Bedroya:2019snp} to relate TCC to bound the exponent in the distance conjecture.}.  Of course there should be other contributions to the potential in order to get a quasi-dS, but because the one-loop term is generically present, the net dependence is still expected to be no weaker than $m^d$.  A higher power of $m$ (i.e. a weaker total potential) would necessitate a magical cancellation of the $m^d$ term.
  Other motivations for the naturalness of the bound $1/d$ on $\alpha$ have been given in \cite{Rudelius:2021oaz,Castellano:2021mmx}.   In this paper, we will thus be assuming the bound in the range of $\alpha$ given in \ref{mainb}.

In the context of the $O(16)\times O(16)$ string the contribution to the energy at weak coupling is positive \cite{Alvarez-Gaume:1986ghj}, because the main contribution comes from the light tower states and it turns out there are more light fermions than bosons and fermions (bosons) contribute positively (negatively) to the one-loop vacuum energy.  As we will see in our universe we also expect an effective potential which falls off at infinity from the positive direction, suggesting that there are more light fermions than bosons in the mass scale of the light tower.
In fact, this observation was already made in \cite{Gonzalo:2021fma}, where it was conjectured that a quasi-dS scenario\footnote{This applies both to dS minima or quintessence models.} consistent with quantum gravity should contain a surplus of light fermions with masses $m\lesssim \Lambda^{1/d}$. This was required to avoid a violation of the non-supersymmetric AdS conjecture \cite{Ooguri:2016pdq} as well as the generalized distance conjecture \cite{Lust:2019zwm} upon further compactification of the theory to lower dimensions.  In this sense, the value $\alpha=1/d$ is special as it directly avoids generating any Casimir vacua in lower dimensions that could be inconsistent with Swampland considerations \cite{Gonzalo:2021fma}.

\section{Swampland criteria and the prediction of the Dark Dimension}\label{sec:dd-3}
As we saw in the last section, Swampland criteria suggest that in extreme limits of parameter space, light KK modes or string excitations emerge.  We now want to apply this to our universe.  One of the extreme parameters in our universe is the cosmological constant which in Planck units is
\begin{equation} \Lambda \sim 10^{-122}.\end{equation}
We now argue that given other observational data in our universe, the smallness of this parameter combined with Swampland principles lead to one extra mesoscopic dimension of length scale of order of a micron.

We start by noticing that a light tower of states causes significant deviations from Newton's law at the energy scale $m$ of the tower. For a KK tower, this is due to gravity propagating on the extra dimension; for the string case, local physics breaks down at the string scale. 
Torsion balance experiments provide the strongest bounds to deviations of the $1/r^2$ gravitational force law. In \cite{Lee:2020zjt}, this law was verified down to the scales around $30\, \mu m$, implying that the scale of the tower has to satisfy,
\beq
m\gtrsim 6.6\, meV.
\eeq

This scale is within an order of magnitude of the energy scale associated to the cosmological constant, $\Lambda^{1/4}=2.31\,meV$ (in terms of length scales, $\Lambda^{-1/4}\approx\, 88\, \mu m$). Hence, the mass scale of the tower must satisfy $m\gtrsim \Lambda^{1/4}$, since otherwise we would have already detected deviations of Newton's law in our universe. The only way to make this experimental bound consistent with the theoretical swampland bound \eqref{alpha} is that this is satisfied with the boundary of the allowed range namely $\frac{1}{d}=\frac{1}{4}$:
\beq
\label{mL}
m\sim \Lambda^{1/4}
\eeq
which coincides with the neutrino scale. We are therefore led to conclude that there is a tower of states starting at the neutrino scale in our universe.

As already mentioned, we expect that this tower is either a light string tower or a light KK tower.
In the first possibility, a perturbative string limit, local EFT breaks down at the string scale $m$. Since we can describe physics above the neutrino scale with effective field theory, the first scenario is ruled out experimentally. Hence, we are left only with the scenario of decompactification.

Assuming this, we must now determine the number $n$ of large dimensions, close to the decompactification limit. Typically, the strongest bounds for the existence of new light weakly coupled particles come from astrophysical data. 
If extra dimensions are present, a new decay mode becomes available through emission of the Kaluza-Klein modes which could leave a trace in neutron stars and supernova explosions. For instance, emission of KK modes
can potentially cool a proto-neutron star too fast to be
compatible with observations if the tower is too light. According to \cite{ParticleDataGroup:2020ssz}, the current strongest bounds come from the heating of old neutron stars by the surrounding cloud of trapped KK gravitons \cite{Hannestad:2003yd,ParticleDataGroup:2020ssz}, yielding
\begin{itemize}
\item For the case of a single extra dimension: $
m^{-1}\sim l < 44\, \mu m 
$
\item For the case of two extra dimensions: $
m^{-1} \sim l <1.6 \cdot 10^{-4}\,\mu m
$
\end{itemize}
For more than two dimensions, the bounds are even more stringent.
This means that the cases of two or more extra dimensions are ruled out in our setup, since they are not compatible with \eqref{mL}. The only possibility is, therefore, that the tower signals decompactification of a single extra dimension, which we call {\it the Dark Dimension}. Using \eqref{MPn}, the model leads to a higher-dimensional Planck scale $\hat{M}\sim\,10^9\, GeV$, a scale at which new physics must come into play.\footnote{Interestingly, using some entropy bounds, the same UV cut-off was pointed out in \cite{Castellano:2021mmx} by assuming an IR cut-off set by the cosmological constant.}

To sum up, we have argued that
\begin{equation}\Lambda \sim  m^{4} \sim 10^{-122} M_{pl}^4.\end{equation}
In other words, we expect $\Lambda^{\frac14}=\lambda \, m$ for some a priori $O(1)$ parameter $\lambda$, and 
where $m$ is related to an inverse length scale $m=l^{-1}$ of one extra mesoscopic dimension. 
 To find $\lambda$ we need extra information.  We will now estimate $\lambda$ and argue that
\begin{equation} m^{1/2}\leq \lambda^4 \lesssim1\quad\Rightarrow\quad \Lambda^{-2/9}\lesssim l\leq\Lambda^{-1/4}.\label{salu2}\end{equation}

\begin{figure}[!htb]
	\begin{center}
		\includegraphics[width=0.6\textwidth]{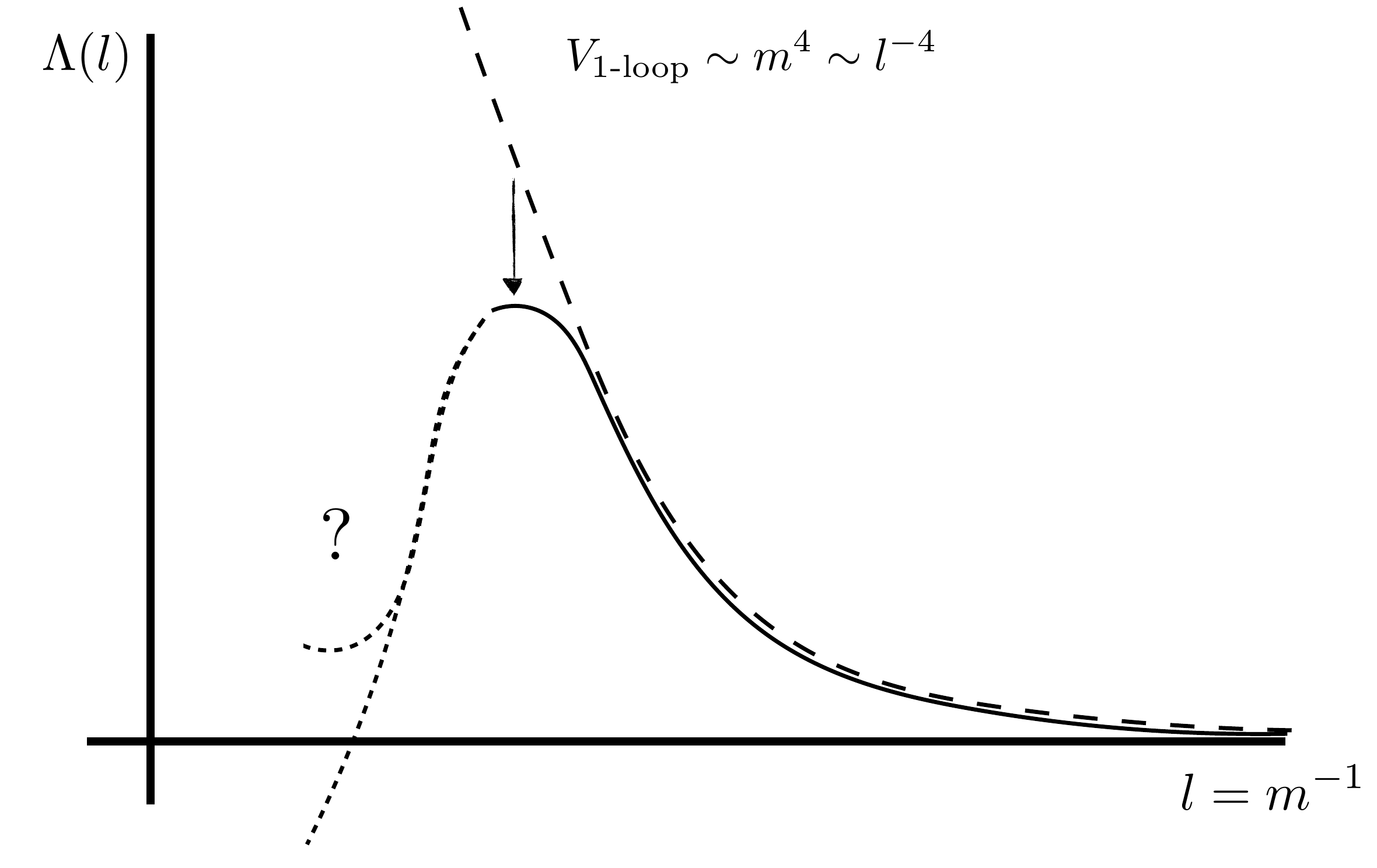}
		\caption{Schematic representation of the vacuum energy of the Dark Dimension as a function of its size. We expect that for large values of the radii, the one-loop effect which goes like $m^4$ dominates the potential leading to excess light fermions. At smaller radii, other effects may take over, resulting in either a (quasi-)dS minimum (dotted lines), or just a maximum. Barring fine-tuning, the location of the maximum is always strictly below the extrapolation of the one-loop vacuum energy curve (as indicated by the arrow), hence yielding $\lambda^4< 1$.}\label{fig1}
	\end{center}
\end{figure}

The lower bound is a consequence of our assumption that the asymptotic limit has already set in:  for $\Lambda\sim \lambda^4 m^4$, the $\lambda^4$ term should not be as small as, say, $m$, for then the scaling would be more appropriately described as $m^5$. $\lambda^4\sim m^{1/2}$ is the lowest it can go while the scaling $\Lambda\sim m^4$ being approximately true. The upper bound is most easily argued for pictorially, as in Figure \ref{fig1}. Since we observe a positive vacuum energy, the potential should approach 0 from above, signalling extra light KK fermions, as noted before.  Moreover the potential should turn down from the $m^4$ pure contribution as that is too steep to give quasi dS. It could then continue as a quintessence model or generate a dS maxima or minima. In any case, from this, it follows that the quasi-static dS always has $\lambda^4<1$.

Another motivation to have $\lambda<1$ comes from the Casimir energy. Recall that the theoretical bound $\alpha\geq 1/d$ in \eqref{mainb} comes from the fact that at the very least, in the absence of supersymmetry, there will be a one-loop Casimir energy contribution to the vacuum energy from the light tower of order $V\sim \lambda_{\rm Casimir}m^4$. This proportionality factor can be explicitly computed for a circle compactification from D=5 to 4 dimensions, yielding
\beq
\lambda_{\rm Casimir}=\frac{\Gamma(D/2)\zeta(D)}{(2\pi)^{D-1}\Gamma(1)\Pi^{D/2}}=\frac{3\zeta(5)}{(2\pi)^6}=5\cdot 10^{-5}
\eeq
Hence, if the correlation $\Lambda^{1/4}=\lambda m$  comes from the Casimir contribution, it is natural to expect a parameter $\lambda\leq \lambda^{1/4}_{\rm Casimir}\sim 10^{-1}$. More concretely, for $D=5$, this value of $\lambda$ leads to 
\begin{equation}l\approx 7.42\, \mu m.\label{nbp}\end{equation}


Note that the correlation between the vacuum energy at a dS maximum and the scale of the tower can be explained for example from a two-term structure of the potential, which can arise in examples of dimensional reduction \cite{Rudelius:2021oaz,Gonzalo:2021fma}. As explored in \cite{Agrawal:2018rcg,Olguin-Trejo:2018zun}, a tachyonic de Sitter scenario is compatible with observational data, and seems potentially easier to embed consistently in quantum gravity (the top of the hill scenario was explored as it is compatible with the dS conjecture \cite{Obied:2018sgi,Garg:2018reu,Ooguri:2018wrx}).  Our analysis is consistent with this scenario but does not require it, as long as we end up with a quasi-dS vacuum.

Thus from \eqref{salu2}, we estimate $10^{-4}<\lambda<1$ or the central region $\lambda\sim 10^{-1}-10^{-3}$,
which leads us to one extra mesoscopic dimension of length scale in the micron range:
\begin{equation}l \sim 0.1\,\mu m\ -\, 10\, \mu m.\end{equation}
The Casimir estimate \eqref{nbp} is in the middle of this range. Note also that this estimate leads to a slightly higher-dimensional Planck scale given by $\hat{M}\sim m^{\frac13} M_{pl}^{\frac23} \sim \lambda^{-\frac13} \Lambda^{\frac{1}{12}}M_{pl}^{\frac23}\sim 10^9-10^{10} GeV$. 

\section{Realization in F-theory}\label{sec:fth-4}

\begin{figure}[!htb]
	\begin{center}
		\includegraphics[width=0.75\textwidth]{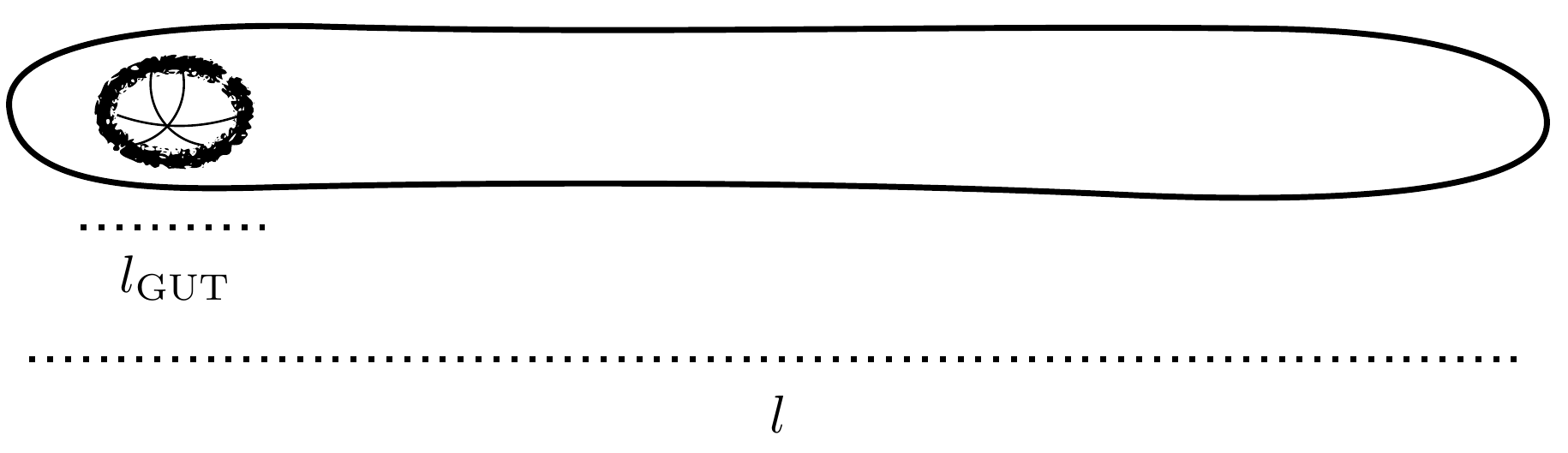}
		\caption{Schematic representation of the Dark Dimension in an F-theory GUT scenario.   Higher towers of SM fields localized in the GUT brane kick in at $\hat{M}$, significantly changing the Higgs potential and preventing the onset of the instability noticed in \cite{Elias-Miro:2011sqh}.}\label{fig2}
	\end{center}
\end{figure}

Swampland ideas leads to the Dark Dimension scenario, but they do not directly tell us how the matter sector is realized. However, if the matter is realized in the bulk of the extra dimension, it leads to too small a coupling.  Therefore it is clear that the matter sector should arise in localized regions of the internal space. A potentially promising realization of this scenario is in the context of F-theory model building and we follow the approach suggested in \cite{Beasley:2008dc,Beasley:2008kw} (see also \cite{Donagi:2008ca}).   Indeed other phenomenological motivations and in particular realization of GUT's already led to the study of this class of models.
In this framework, grand unification occurs on a GUT 7-brane which is a contractible cycle in the base of a Calabi-Yau 4-fold.  The volume of the 7-brane in the higher dimensional Planck units is fixed by the coupling constant of the standard matter fields, 
\begin{equation} \frac{1}{e^2}=V_{\text{SM}}\,\hat{M}^4\sim 10^2,\end{equation}
leading to a size for the GUT brane where the Standard Model fields reside $l_{std}\sim O(1) {\hat M}^{-1}$ (see Figure \ref{fig2}).  In particular this leads to tower of matter charged under standard model excitations, coming from the tower of states of the 7-brane, at a scale $\sim 1/l_{Std}\sim {\hat M}\sim 10^{9}-10^{10}GeV$. It is natural to expect this scale to coincide with the SUSY breaking scale (at least it may be an upper bound for it).  In such a case the Higgs potential, which is otherwise expected to exhibit instability around $10^{11}GeV$ \cite{Elias-Miro:2011sqh}, will be expected to have its stability restored at higher energies due to corrections from the tower of SM fields as well as the restoration of supersymmetry, as explored e.g. in \cite{Ibanez:2013gf}.  Of course, at this mass scale we also encounter the higher dimensional black holes. All this suggests that we do not need to worry about instability of the Higgs vacuum since the running is surely going to change the potential before the instability arises. 

One may naturally ask how this story fits with the GUT scale in the $10^{15}-10^{16}GeV$ range expected from low energy considerations, which is above the five-dimensional Planck scale.
From an EFT point of view, it is natural to expect that strong threshold corrections (see e.g. \cite{Antoniadis:1990ew,Conlon:2009qa,Ibanez:2012zg}) will dramatically modify the running of the couplings. This however does not mean that the GUT scale needs to be brought down to $\hat{M}$, as the GUT scale is determined by hypercharge flux in F-theory models which can be large\footnote{Note that the flux cannot be much larger than $10^4-10^5$ (bounded by tadpole conditions) \cite{Becker:1996gj}.}.  This leads to GUT scales which can be higher by a few orders of magnitude.
  It is important to explore various options of GUT realization and the related issue of proton stability further in this scenario\footnote{For example, in \cite{Ibanez:2012zg} it is argued that the hypercharge flux deforms the SM fermion wave functions leading to a suppression, avoiding in this way the strong experimental proton decay constraints.}.


Also the issue of realizing a quasi-dS background in this setup is important.  In this connection one is naturally led to large volume scenarios in the context of F-theory which has been studied in \cite{Balasubramanian:2005zx}.  Unfortunately the technical issues related to obtaining dS or even quasi-dS solutions in this context are still not settled (see e.g.\cite{Junghans:2022kxg,Junghans:2022exo}).

We also need to consider the couplings between the scalar parametrizing the size of the Dark Dimension (radion) and the matter sector, to avoid trouble with fifth-force constraints \cite{Lee:2020zjt}. One way to do this is to make the radion sufficiently heavy; another is to suppress the couplings of the radion to the matter sector which will be localised on the GUT branes, and could depend e.g. on the geometry of additional compact dimensions. This is an interesting question that should be explored further in the future.

\section{Further phenomenological predictions of the dark dimension scenario}\label{sec:pheno-5}
The basic observation of this paper is that, if we entertain the possibility that our universe is close to an infinite distance limit, experimental considerations only allow this to be a mesoscopic extra dimension, which is tied together with the vacuum energy. As a result, several phenomena typically associated to the large extra dimensions scenario  \cite{Arkani-Hamed:1998jmv,Arkani-Hamed:1998wuz,Dienes:1998sb} appear for us. The main difference is that the predicted scales are completely different, reflecting the different motivations between the two scenarios.
Perhaps the most direct way to check our scenario is to sharpen the impressive experiments \cite{Lee:2020zjt} testing deviations from the Newton's gravitational force law to an even higher precision.  Indeed the current precision is at the doorstep of our prediction and a factor of $10-100$ improvement in the precision of this experiment would bring us to a regime to seriously test our scenario.  There are other phenomenological implication of our scenario and in this last section we review a few of these implications focusing on the new connections that follow from our Swampland principles.

\subsection*{Neutrino masses and $H_0$ tension}
In a large extra dimension, the towers that become light are KK modes.  Since $V\sim m_{KK}^4$, and experimentally $V\sim m_\nu^4$ is close to the neutrino mass scale, it is natural to identify the tower with a tower of sterile, right-handed neutrinos, as done in \cite{Dienes:1998sb,Arkani-Hamed:1998wuz}\footnote{This identification is also useful to avoid the generation of Casimir vacua that would violate the Swampland conjectures upon compactifying the Standard Model in a circle \cite{Ooguri:2017njy,Gonzalo:2018dxi,Gonzalo:2021zsp}.}. Note that this is particularly natural in our setup, because as we have already discussed the sign of the vacuum energy suggests that we have an excess of light fermions some of which could be viewed as giving rise to this tower.
Moreover, as described in section \ref{sec:fth-4}, we do not expect to find a tower with the quantum numbers of the SM until the species scale $\hat{M}$. Therefore we expect light left-handed neutrinos on the GUT brane that couple to a KK tower of right-handed ones in the bulk.  Precisely this scenario is explored exhaustively in \cite{Carena:2017qhd}, from the bottom-up point of view and due to its phenomenological appeal. As explained there, schematically there is an effective $2\times 2$ mass matrix for each flavor \cite{Dienes:1998sb,Arkani-Hamed:1998wuz,Carena:2017qhd} of the form  
\begin{equation}\mathcal{M}=\left(\begin{array}{cc} 0&\frac{y \langle H\rangle}{\sqrt{ l\hat{M}}}\\
 \frac{y \langle H\rangle}{\sqrt{ l\hat{M}}}&\frac{1}{l}
\end{array}\right),\label{matrix}\end{equation}
with $l$ being the size of the extra dimension, as before, and $y$ being the Yukawa coupling of the GUT brane with the bulk states.  The $1/{\sqrt l}$ comes from the wave function normalization of the bulk 5d fermions.  Assuming $1/l$ is bigger than or equal to the off diagonal term we can estimate the lower eigenvalue corresponding to the active neutrino to be
\begin{equation} m_\nu\approx   \frac{y^2\langle H\rangle^2}{\hat{M}},\end{equation}
modulo $\mathcal{O}(1)$ factors that are controlled by the effective number of states in the tower, and can be tuned by turning on a bulk mass for the neutrino, if necessary. 
This is an expression similar to the usual see-saw mechanism, with $\hat{M}\approx\, 10^9-10^{10}\, GeV $ playing the role usually taken by $M_{\text{GUT}}$. A Yukawa coupling of order $y\sim10^{-2}-10^{-3}$ is enough to bridge this gap and produce the correct values for the neutrino masses.

 Clearly, there are many more detailed questions one needs to ask in this scenario. For instance, the additional neutrino oscillations caused by the tower of sterile neutrinos must be made compatible with experimental constraints. The work \cite{Carena:2017qhd} shows that this is at least possible. Another concern is the production of the tower of neutrinos at LHC, for instance via Higgs decays. Taking again results from \cite{Carena:2017qhd}, the decay rate of Higgs to towers of KK modes can be estimated as
 \begin{equation}\Gamma(h\,\rightarrow\,\text{Tower})\approx \frac{M^2_hy^2}{\hat{M}}\sim m_\nu,\end{equation}
 which is small and easily avoids all the experimental bounds. Notice how, throughout all of this discussion, there is a single mass scale, the mass scale of the tower (or equivalently the dark energy), which controls all the relevant aspects of the physics.

Additional neutrino species may also be in trouble with cosmology, since these can provide additional contributions to $N_{\text{eff}}$, the number of relativistic degrees of freedom, which is very constrained after electron-positron annihilation \cite{Carena:2017qhd,Hagstotz:2020ukm} to be close to 3. In order to avoid these constraints, the mixing between sterile and active neutrinos must be small enough so that they are not significantly populated, providing a further constraint on the model.  As will be discussed in \cite{nextpaper} the initial temperature can be in the TeV range, without the KK excitations causing an issue for the $N_{\text{eff}}$, thus avoiding any modification of the BBN (see also \cite{Arkani-Hamed:1998wuz}).

An extra-dimensional model can also potentially provide an explanation of the $H_0$ tension \cite{DiValentino:2021izs}, along the lines of \cite{Sakstein:2019fmf,CarrilloGonzalez:2020oac}. In this reference, it is pointed out that the $H_0$ tension can be significantly alleviated if there is a component of dark energy that becomes active around the time of matter-radiation equality (known as ``Early Dark Energy"), and furthermore that a natural scenario is to introduce a new scalar that couples to the neutrino mass term. This coupling would induce a kick to the energy density of the universe when the temperature of the thermal background is around the neutrino mass scale.  However it is difficult to justify, from a theoretical point of view, why it has a coupling to neutrino mass terms. Both questions are naturally accommodated in our scenario if we interpret the scalar to be the ``radion'' field of the extra dimension; the coupling to neutrinos arises simply from the fact that the tower of KK modes is coupled to the 5d metric and thus controls the neutrino masses by the seesaw mechanism reviewed here. 

It is imperative to perform a quantitative analysis of the possible behaviors that a neutrino tower can have according to the Distance Conjecture, and obtaining universal predictions for neutrino masses and oscillations. Similarly, the scenario we just outlined for the $H_0$ tension needs to be taken beyond the qualitative level. We will pursue these directions and more generally the cosmological aspects of this scenario in a future publication \cite{nextpaper}.

\subsection*{Unification of hierarchies} 

The identification of the asymptotic tower with sterile right-handed neutrinos also allows to provide a correlation between the EW hierarchy problem and the cosmological constant problem, as we now explain. The mass matrix \eqref{matrix} yields two different scales for the mass eigenstates,
\begin{itemize}
    \item $m_\nu\sim \frac{y^2\langle H\rangle^2}{\hat M}$ for the lightest (mostly active) neutrinos
    \item $M_\nu \sim \frac{1}{l}$ for the heavier sterile neutrinos
\end{itemize}
In principle the eigenvalues of the neutrino mass matrix can have vastly different scales. Assuming no unnatural mass hierarchies in the neutrino sector, leads to relating the weak scale hierarchy to the dark energy.  Namely for the active and sterile neutrinos to have similar mass scales\footnote{Neutrino oscillations can be sensitive to the exact mass splitting between active and sterile neutrinos \cite{Carena:2017qhd}, a topic we hope to return to in the future \cite{nextpaper}.} means that the mechanism for stabilization of this dark dimension also sets the scale for the Higgs potential.  We will not offer specific mechanisms for this in the present paper, but hope to return to it in a future work.  With this assumption, we are led to the following prediction for the Higgs vev, 
\begin{equation}
\label{Higgs}
    \frac{y^2\langle H\rangle^2}{\hat M}\sim \frac{1}{l}\Rightarrow \langle H\rangle \sim \frac{1}{y}\sqrt{\frac{\hat{M}}{l}}\sim \frac{\Lambda^{1/6}M_{pl}^{1/3}}{y\, \lambda^{2/3}} \sim 10-10^3\, GeV
\end{equation}
where we have used that $\hat M\sim l^{-1/3}M_{pl}^{2/3}$ and $l=\lambda\,\Lambda^{-1/4}$ with $\lambda\sim 10^{-1}-10^{-3}$, $y\sim 10^{-2}-10^{-3}$. Interestingly, this is not far from the experimental result for the Higgs vev\footnote{Experimentally, the active neutrino masses are upper bounded by $\sum_{i=1}^3 m_{\nu_i}\leq 0.139(0.174)$ eV for NH(IH) neutrinos according to \cite{GAMBITCosmologyWorkgroup:2020rmf} (or $\sum_{i=1}^3 m_{\nu_i}\leq 0.09$ eV according to the most recent analysis \cite{DiValentino:2021hoh}). 
This implies that $\sum_i m_{\nu_i}< \Lambda^{1/4}\sim M_\nu$, so that the the above result \eqref{Higgs} which was obtained by setting the masses of active and sterile neutrinos equal is a slight over-estimation of the Higgs vev.}.  Note that in our scenario we have tied all the questions of hierarchy to the dark energy.  Namely we have conntected the Higgs vev $\langle H\rangle  \sim y^{-1}\lambda^{-2/3} \Lambda^{1/6} M_{pl}^{1/3}$, the mesoscopic length scale and neutrino masses $l\sim  \lambda \Lambda ^{-1/4}\sim 1/m_\nu$, as well as a new UV scale in physics ${\hat M}\sim \lambda^{-1/3} \Lambda^{1/12}M_{pl}^{2/3}$ all to the dark energy\footnote{For another way the logic of naturalness in EFTs must be modified in the presence of gravity see \cite{Ibanez:2017kvh}.}.
Typically, such small values for the Higgs vev seem unnatural from a bottom-up perspective. However, when identifying the tower of states with sterile neutrinos, the smalleness of the Higgs vev can emerge from the smallness of the cosmological constant, so that the two hierarchy problems are reduced to a single one. Of course, one still needs to explain why the cosmological constant is small, which we relate to the fact that we live near an asymptotic boundary of the field space, i.e. a large field distance limit.

\subsection*{Ultra-high energy cosmic rays and the GZK limit}
Astrophysical sources can accelerate particles to energies much beyond those that can be probed by LHC, generating ultra-high energy cosmic rays that reach the Earth and produce showers that can be detected \cite{Anchordoqui:2018qom}. Current predictions for cosmic ray abundance typically assume that ultra-high energy cosmic rays are mostly comprised by heavy nuclei and other charged particles accelerated to enormously high energies by magnetic fields \cite{Anchordoqui:2018qom,AlvesBatista:2019tlv}. With this assumption, the spectrum of ultra-high energy cosmic rays can be reasonably explained, including somewhat detailed features, up to energies of around $10^9-10^{10}$ GeV, where the abundance of cosmic rays suddenly drops exponentially. One simple way to explain this suppression is that at these energy scales, scattering between the high-energy cosmic rays and CMB photons becomes efficient, effectively turning the intergalactic medium into an opaque material for the ultra-high energy cosmic rays. This upper bound in energy is known as the GZK limit \cite{Greisen:1966jv,Zatsepin:1966jv}, and the theoretical value roughly coincides with the observed supression in the cosmic ray spectrum.  It is however still an open question whether the sharp cutoff observed in the data at very high energies is indeed the GZK mechanism or has to do with intrinsic limitations of the sources of cosmic rays \cite{AlvesBatista:2019tlv}.

The GZK energy scale of roughly $10^9-10^{10}$ GeV coincides with the species scale/ 5d Planck mass scale in our scenario. As a result, we would expect a qualitative change in the behavior of cosmic rays starting at around this energy scale. Particles with masses above this scale necessarily have internal structure in the fifth dimension, and scattering of these would lead to a significant production of KK modes. Localized objects in the fifth dimension necessarily become black holes at this energy scale. While it is difficult to be precise, our scenario would predict significant deviations in the spectrum of ultra-high energy cosmic rays beyond the GZK limit. 

It is frustrating that the GZK scale coincides with $\hat{M}$, as that prevents us from checking our scenario directly. On the bright side, the GZK limit does not imply that there cannot be sources of cosmic rays with energy beyond $10^9$ GeV; it only says cosmic rays of higher energies which are produced sufficiently far away enough (beyond the so-called GZK sphere, of radius  $50\, Mpc$) will not reach us. Higher sensitivity to ultra-high energy cosmic rays in our vicinity may make some of the effects we discuss observable. Discrimination of ultra-high energy cosmic ray sources within and outside the GZK sphere is one of the current experimental priorities \cite{AlvesBatista:2019tlv}.

\section*{Acknowledgments}

We have greatly benefited from discussions with Prateek Agrawal, Nima Arkani-Hamed, Greg Gabadadze, Eduardo Gonzalo, Jonathan Heckman, Luis Ib\'{a}ñez, Georges Obied, Eran Palti, Matthew Reece, Tom Rudelius, Iv\'{a}n Mart\'{i}nez Soler and Mark Trodden.
The work
of MM and CV is supported by a grant from the Simons Foundation (602883, CV) and by the
NSF grant PHY-2013858. The work of IV is partly supported by the grants RYC2019-028512-I from the MCI (Spain), PGC2018-095976-B-C21 from MCIU/AEI/FEDER, UE and the grant IFT Centro de Excelencia Severo Ochoa SEV-2016-0597.

\bibliographystyle{JHEP}
\bibliography{refs-darkdim}

\end{document}